# Backbone chemical shift assignments of human 14-3-3σ

João Filipe Neves[1], Isabelle Landrieu[1]*, Hamida Merzougui[1], Emmanuelle Boll[1], Xavier Hanoulle[1], François-Xavier Cantrelle[1]

[1]UMR 8576 CNRS-Lille University, 59000 Lille, France

ORCID ID for authors:
JFN: 0000-0003-0582-1193
IL: 0000-0002-4883-2637
XH: 0000-0002-3755-2680

*Corresponding author: Isabelle Landrieu, isabelle.landrieu@univ-lille1.fr
+33 (0)3 62 53 17 02

**Abstract**

14-3-3 proteins are a group of seven dimeric adapter proteins that exert their biological function by interacting with hundreds of phosphorylated proteins, thus influencing their sub-cellular localization, activity or stability in the cell. Due to this remarkable interaction network, 14-3-3 proteins have been associated with several pathologies and the protein-protein interactions established with a number of partners are now considered promising drug targets. The activity of 14-3-3 proteins is often isoform specific and to our knowledge only one out of seven isoforms, 14-3-3ζ, has been assigned. Despite the availability of the crystal structures of all seven isoforms of 14-3-3, the additional NMR assignments of 14-3-3 proteins are important for both biological mechanism studies and chemical biology approaches. Herein, we present a robust backbone assignment of 14-3-3σ, which will allow advances in the discovery of potential therapeutic compounds. This assignment is now being applied to the discovery of both inhibitors and stabilizers of 14-3-3 protein-protein interactions.

**Biological context**

The 14-3-3 family includes seven different isoforms (β, γ, ε, η, σ, τ and ζ) expressed in all eukaryotic organisms. 14-3-3 proteins are adapter proteins, expressed in an ubiquitous manner in all tissues and cell compartments. Despite the considerably high structural homology between different isoforms, 14-3-3 activity is often isoform specific (Fu et al. 2000). To our knowledge, 14-3-3ζ is the only isoform of 14-3-3 whose assignment is reported (Killoran et al. 2015). 14-3-3 proteins exert their biological functions through the modulation of the activity of hundreds of phosphorylated proteins. This remarkable interactome makes 14-3-3 proteins influent actors in many cellular events and, by consequence, in several pathologies like cancer, Alzheimer's disease, Parkinson's disease, among others (Kaplan et al. 2017) . With the rise of the interest on the modulation of Protein-Protein Interactions (PPIs) in modern drug discovery, 14-3-3 proteins are being seen as promising targets for several pathologies. Especially over the last 10 years, there has been a remarkable effort towards the discovery of compounds capable of inhibiting or stabilizing 14-3-3 PPIs (Stevers et al. 2018). In particular, we have been interested in modulating the Tau/14-3-3 interaction, which is thought to have detrimental effects in neuronal cells in Alzheimer's disease (Joo et al. 2015). Peptidomimetic compounds have been designed and shown to inhibit this PPI (Milroy et al. 2015; Andrei et al. 2018). Assignment of 14-3-3σ spectra will further help the development of these promising compounds.

In physiological conditions, 14-3-3 proteins exist as functional dimers, both as hetero and homodimers (Fu et al. 2000). They are acidic proteins and each monomer has a molecular weight (MW) of around 30 kDa, being composed by nine α-helices, commonly named αA-I, disposed in an antiparallel fashion. At the C-terminus of the last helix of 14-3-3 proteins (αI), there is a flexible and highly acidic tail, which is the less conserved region of 14-3-3 (Obsil and Obsilova 2011).

Here we report the backbone assignments ($^{13}C_\alpha$, $^{13}C_\beta$, $^{13}CO$, $^{15}N$ and $^{1}H^N$) of 14-3-3σ (a truncated form comprising residues 1-231 out of 248). Due to the large number of partners of 14-3-3σ, the assignment of this protein opens new possibilities for the study of hundreds of PPIs. Our assignment is currently being applied to drug screening and to the study of the effect of small molecules on the modulation of 14-3-3 PPIs. In the future, it may have a role on the development of new 14-3-3 PPI modulators that can be employed as tool compounds and hopefully, be the basis for the development of new therapeutic molecules.

**Sample preparation**

The plasmids used for protein expression were kindly provided by Dr. Christian Ottmann from the Eindhoven University of Technology, The Netherlands. The cDNA coding for human 14-3-3σ (residues 1-248) or 14-3-3σΔC17 (residues 1-231) was cloned into the pProExHtb vector between the *Bam*HI and *Not*I sites. A Tobacco Etch Virus (TEV) cleavage site followed by a linker (GAMGS) are present between the regions coding for the N-terminal His$_6$-tag and for the protein, yielding a recombinant expression plasmid with the following configuration: His$_6$ tag-TEV cleavage site-Linker-14-3-3σ/14-3-3σΔC17. Therefore, 14-3-3σ and 14-3-3σΔC17 were both expressed as N-terminally His$_6$-tagged proteins.

$^{15}$N$^{13}$C$^{2}$H labeled 14-3-3σ or 14-3-3σΔC17 were expressed in *E. coli* BL21 (DE3) cells. A 20 ml pre-culture in *Luria–Bertani* (LB) medium containing 100 mg/L ampicillin was grown overnight at 310 K and was used to inoculate 1 L of deuterated M9 minimal medium supplemented with 2 g/L $^{13}$C$_6$$^{2}$H$_7$ D-Glucose, 1 g/L $^{15}$N Ammonium Chloride, 0.4 g/L Isogro $^{15}$N$^{13}$C$^{2}$H Powder – Growth Medium (Sigma Aldrich) and 100 mg/L ampicillin. The culture was grown at 310 K to an OD$_{600}$ of 0.9 and induced with 0.5 mM Isopropyl-β-D-Thiogalactopyranoside (IPTG). Incubation was continued for 15h at 301 K with vigorous shaking. Cells were harvested by centrifugation, resuspended in a buffer containing 50 mM Tris-HCl, 500 mM NaCl, 5% (v/v) glycerol, 25 mM imidazole, 2 mM β-Mercaptoethanol (BME), pH 8.0 and lysed with a homogenizer. The cellular debris were then eliminated by centrifugation and the supernatant was loaded into a Ni-NTA column (GE Healthcare). Elution of proteins was performed with a buffer containing 50 mM Tris-HCl, 500 mM NaCl, 5% (v/v) glycerol, 250 mM imidazole, 2 mM BME, pH 8.0. The N-terminal His$_6$-tag was then cleaved by the TEV protease for 2 hours at 293 K, followed by 12 hours at 277 K, while being dialyzed against 100 mM sodium phosphate, pH 6.8 and 50 mM NaCl (NMR buffer). The molar ratio between 14-3-3σ or 14-3-3σΔC17 and TEV protease was 50:1. The protein without His$_6$-tag was collected in the flow-through of a Ni-NTA column, using NMR buffer for the elution, while the uncleaved fraction and the His$_6$-tagged peptide were retained on the column. The protein was then concentrated to 2 mM, aliquoted, flash frozen and stored at 193 K. Typical yields were in the range of 45-70 mg of protein per liter of culture.

**NMR spectroscopy**

All NMR experiments were recorded using a 900 MHz Bruker Avance spectrometer, equipped with a cryoprobe. Samples, in a buffer containing 100 mM sodium phosphate, pH 6.8, 50 mM NaCl, 1mM DTT (Dithiothreitol), EDTA-free Protease Inhibitor Cocktail (Roche, Switzerland) and 10% (v/v) $D_2O$, were transferred to *Shigemi* tubes and experiments were acquired at 305 K. The concentration of $^{15}N^{13}C^{2}H$ labeled protein for assignment experiments ranged from 0.8 mM to 1.0 mM. Backbone assignments were obtained from TROSY-HNCACB, TROSY-HN(CO)CACB, BEST-TROSY-HNCO and $^{1}H$-$^{15}N$-NOESY-HMQC spectra on $^{15}N^{13}C^{2}H$ labeled 14-3-3σ or 14-3-3σΔC17 samples. The mixing time of the $^{1}H$-$^{15}N$-NOESY-HMQC spectrum was 120 ms. Non-Uniform Sampling was employed for the acquisition of all 3D spectra. The reference for the $^{1}H$ chemical shift was relative to Trimethyl silyl propionate. The $^{15}N$ and $^{13}C$ chemical shifts were referenced indirectly. All spectra were collected and processed with Topspin 3.5 (Bruker Biospin) and analyzed with Sparky 3.12 (T. D. Goddard and D. G. Kneller, SPARKY 3, University of California, San Francisco). Automatic assignments were performed with MARS version 1.2 (Jung and Zweckstetter 2004) and with PINE-NMR server (Bahrami et al. 2009).

**Assignments and data deposition**

For sequence numbering purposes, the residues of the linker left after TEV cleavage (GAMGS) were not considered. Accordingly, residue 1 corresponds to the first Methionine residue of 14-3-3σ. Similarly to what has been reported for the assignment of 14-3-3ζ (Killoran et al. 2015), we also observed the presence of highly intense peaks in the $^{1}H$-$^{15}N$ TROSY-HSQC spectrum of the full-length 14-3-3σ (residues 1-248). Because these intense signals were overlapping with signals corresponding to the ordered core-region, we decided to use a construct of 14-3-3σ without the C-terminal flexible region (14-3-3σΔC17, containing residues 1-231) for the backbone assignment. The elimination of the resonances corresponding to this flexible region indeed allowed to detect more resonances on the $^{1}H$-$^{15}N$ TROSY-HSQC spectrum and additionally improved the overall quality of the 3D spectra. Removal of the disordered terminal tail did not affect the structure of the protein in solution as resonances in the 2D spectra of the protein with or without the tail superimposed. The final construct, after cleavage of the N-terminus by the TEV protease, yielded a 14-3-3σΔC17 protein with a MW of 26.51 KDa. Given its association as a dimer and therefore the formation of a biomolecule with a MW higher than 50 KDa, it was necessary to produce a deuterated protein, to use a high-field spectrometer and to extend acquisition times. A good coverage of the protein assignment was reached although it should be noted that not all the amide peaks of the protein were detected on the $^{1}H$-$^{15}N$ TROSY-HSQC (shown in **figure 1 a**) and in some cases, backbone signals in the $^{13}C$ dimension were of low intensity. The incomplete $^2$H/$^1$H exchange may have been a factor that prevented the detection of some resonances, in particular those corresponding to residues located at the dimer interface.

Overall, we were able to obtain a robust assignment of 14-3-3σΔC17 comprising 177 out of 226 $^1$H$^N$ and $^{15}$N resonances (78%), 200 out of 231 $^{13}$C$_α$ resonances (87%), 187 out of 219 $^{13}$C$_β$ resonances (85%) and 176 out of 231 $^{13}$CO resonances (77%). The assignment covers homogeneously the protein's sequence (**figure 1 b**). The assigned chemical shifts were deposited into the Biological Magnetic Resonance Database (http://www.bmrb.wisc.edu/) under the BMRB accession number 27563.

**Secondary structure**

The obtained chemical shift assignments were further used for the analysis of the secondary structure of 14-3-3σ by two different methods. On one hand, the ensemble of backbone chemical shifts ($^{13}$C$_α$, $^{13}$C$_β$, $^{13}$CO, $^{15}$N and $^1$H$^N$) was submitted to CSI 3.0 sever (Hafsa et al. 2015) in order to obtain the Chemical Shift Indexes predictive of the secondary structure. On the other hand, the $^{13}$C$_α$ and $^{13}$C$_β$ chemical shift values were used to calculate the Secondary Structure Propensity (SSP) scores (Marsh et al. 2006). Prediction by SSP scores allowed comparison with the secondary structure analysis of the previously assigned 14-3-3 isoform (Killoran et al. 2015).

The results of the predictions made by both softwares are summarized in **figure 2**. Indeed, the results obtained from both methods are coherent and well correlated with what is reported in the crystal structure of 14-3-3σ (PDB ID: 1YZ5). The SSP scores of 14-3-3σ presented here are almost identical to the ones of 14-3-3ζ showing that the secondary structures of both isoforms in solution are very similar. The only remarkable difference is the length of helix αD, which is shorter in 14-3-3σ at its N-terminus due to the replacement of Q$_{79}$ of 14-3-3ζ by a proline residue.

**Acknowledgments**

We thank MSc. Eline Sijbesma and Dr. Christian Ottmann from the Eindhoven University of Technology for kindly providing us the plasmids of both 14-3-3σ and 14-3-3σΔC17. We also thank Dr. Elian Dupré from Lille University for his help with the automatic assignments software.

The research is supported by funding from the European Union through the TASPPI project (H2020-MSCA-ITN-2015, grant number 675179) and by the LabEx (Laboratory of Excellence) DISTALZ (ANR, ANR-11-LABX- 009). The NMR facilities were funded by the Nord Region Council, CNRS, Institut Pasteur de Lille, the European Community (ERDF), the French Ministry of Research and the University of Lille and by the CTRL CPER cofunded by the European Union with the European Regional Development Fund (ERDF), by the Hauts de France Regional Council (contract n°17003781), Métropole Européenne de Lille (contract n°2016_ESR_05), and French State (contract n°2017-R3-CTRL-Phase 1). We acknowledge support for the NMR facilities from TGE RMN THC (CNRS, FR-3050) and FRABio (Univ. Lille, CNRS, FR-3688).

**Conflict of interest**

The authors declare that they have no conflict of interest.

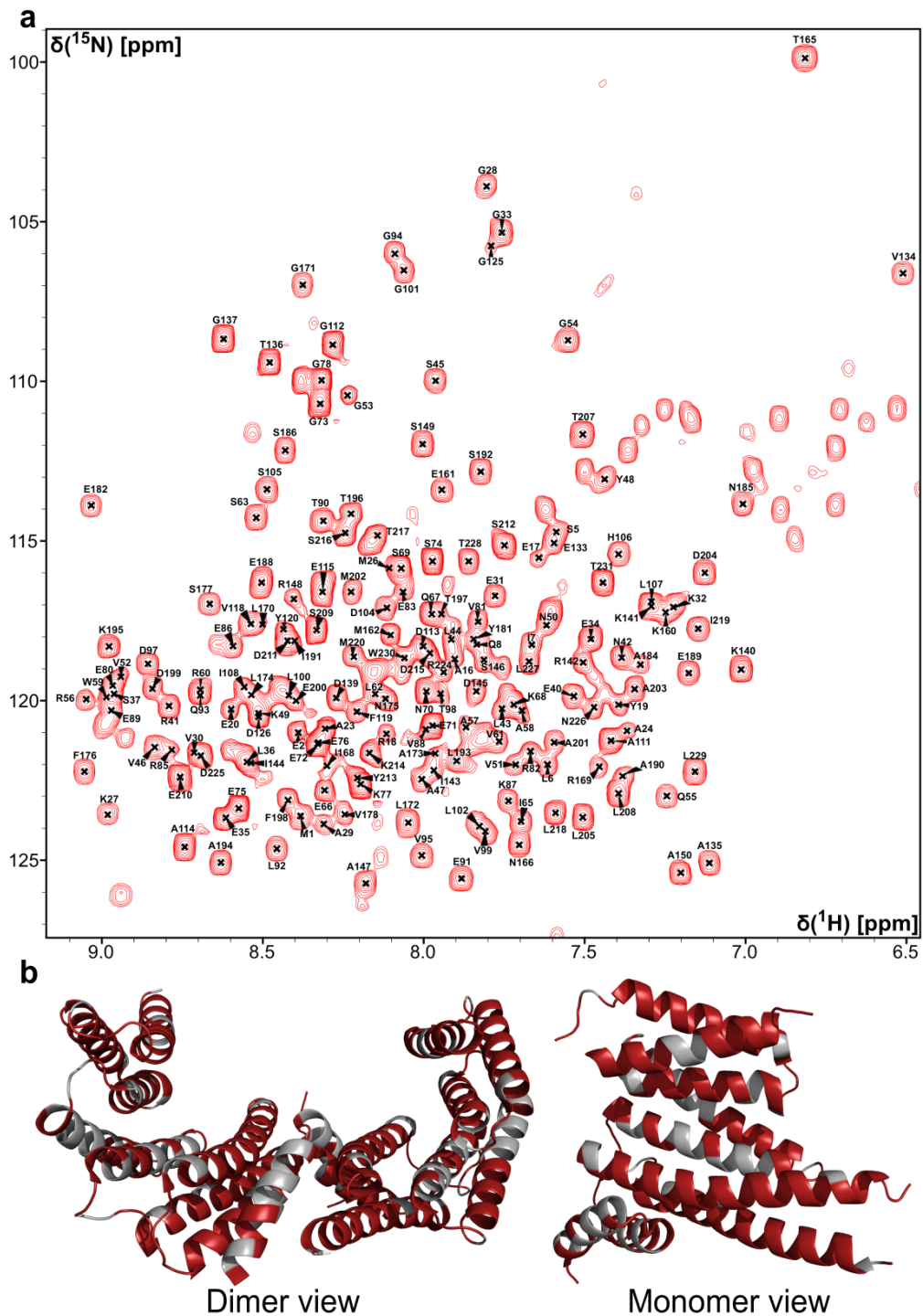

**Fig. 1** Assignment of 14-3-3σ is robust and covers homogeneously the protein's surface. a) Annotated $^1$H-$^{15}$N TROSY-HSQC spectrum of $^{15}$N$^{13}$C$^2$H labeled 14-3-3σΔC17 0.8 mM acquired at 305 K in a buffer containing 100 mM sodium phosphate, pH 6.8, 50 mM NaCl, 1mM DTT, EDTA-free Protease Inhibitor Cocktail (Roche, Switzerland) and 10% (v/v) D$_2$O. The spectrum was acquired with 3426 complex points in the $^1$H dimension and 256 complex points in the $^{15}$N dimension with 64 scans per increment. The spectral window was between 4.72 ppm and 14.64 ppm for the $^1$H dimension and between 80.2 ppm and 140.1 ppm for the $^{15}$N dimension. The selected window contains all the assigned backbone resonances. b) Cartoon representation of the crystal structure of 14-3-3σ (PDB ID: 1YZ5, gray cartoon) with the assigned residues colored in red. The structure is shown both as a dimer and as a monomer of 14-3-3σ.

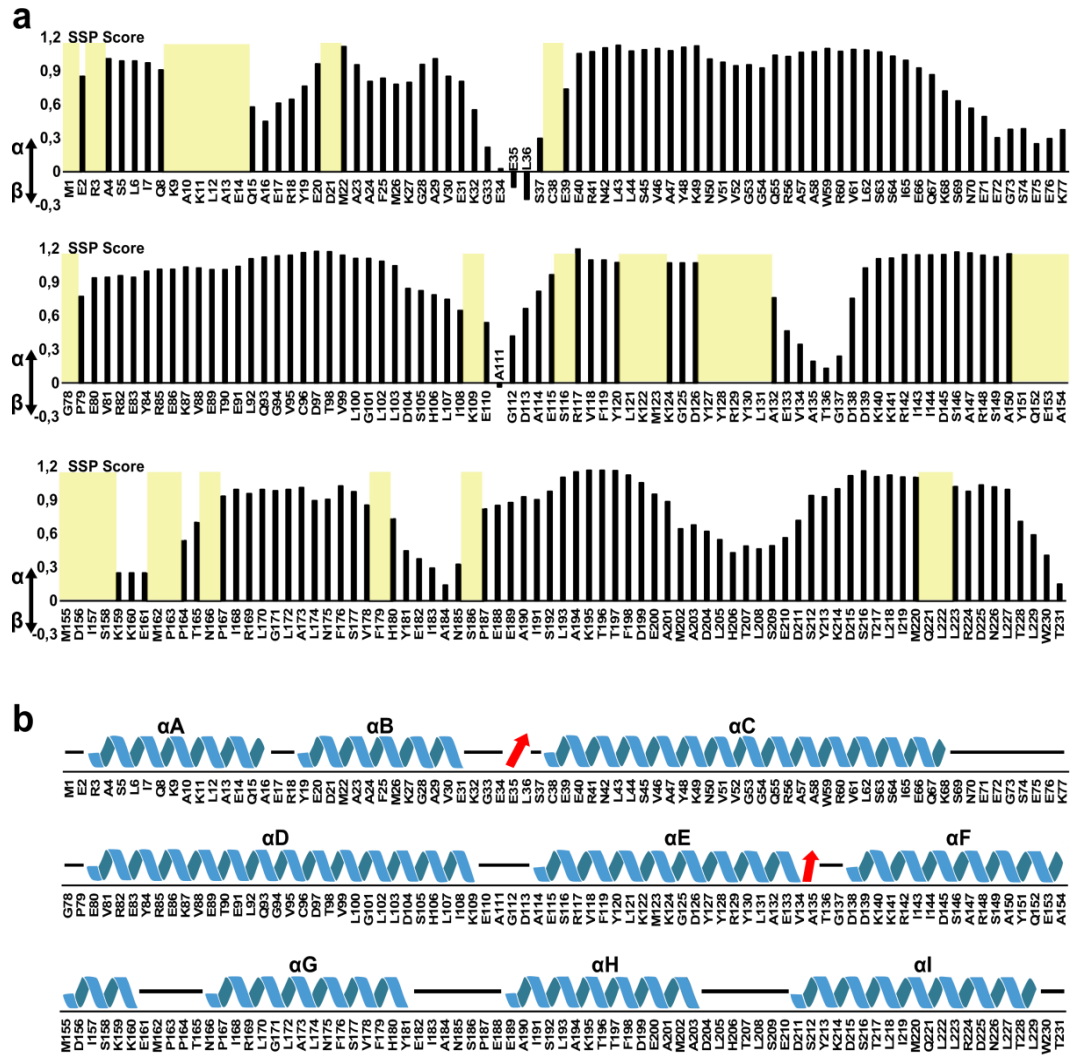

**Fig. 2** Secondary Structure analysis of 14-3-3σ by calculation of the SSP scores and by the CSI 3.0 server reveals that 14-3-3σ is composed by nine α-helices, as expected from the crystal structure. a) SSP scores calculation for 14-3-3σ. The *y axis* of the plot represents the score (~ 1 for α-helix, ~ 0 for random coil, ~ -1 for β-strand) for each residue of 14-3-3σ (represented on the *x axis*). Yellow bars correspond to non-assigned residues or to residues preceding prolines, which were not considered for the SSP score calculation. Random coil chemical shift values used for the calculation of SSP scores were taken from RefDB database (Zhang et al. 2003). The plot is divided in 3 lines. b) Output of the CSI 3.0 server in a residue specific-way (blue cartoon stands for α-helix, red arrow stands for β-strand and black line stands for random coil). The plot is divided in 3 lines. The identification of the helices of 14-3-3σ (αA-αI) is shown above the output of the CSI 3.0 server.